\documentclass[11pt]{article}
\usepackage{epsfig}

\begin{document}

%%%%%%%%%%%%%%%%%%%%%%%%%%%%%%%%%%%%%%%%%%%%%%%%%%%%%%%%%%%%%%%%%%%%%%%%%%%%%
% from the dvips manual: put a background `DRAFT' on the page
%  \special{!userdict begin 
%  /bop-hook{gsave 200 30 translate 65 rotate
%           /Times-Roman findfont 216 scalefont setfont
%           0 0 moveto 0.95 setgray (DRAFT) show grestore}def end}
%
%%%%%%%%%%%%%%%%%%%%%%%%%%%%%%%%%%%%%%%%%%%%%%%%%%%%%%%%%%%%%%%%%%%%%%%%%%%%%

\title{Why does the meter beat the second?}
\author{P.~Agnoli$^1$ and G.~D'Agostini$^2$\\
\mbox{} \\
{\small $^1$~ELECOM s.c.r.l, R\&D Department, Rome, Italy} \\
{\small (paolo.agnoli@fastwebnet.it)} \\
{\small $^2$~Universit\`a ``La Sapienza'' and INFN, Rome, Italy} \\
{\small (giulio.dagostini@roma1.infn.it, 
www.roma1.infn.it/\,$\tilde{ }$\,dagos)}
}

\date{ }

\maketitle
%\vspace{0.3cm}
\begin{abstract}
The French Academy units of  time and length
--- the second and the meter ---
are traditionally considered independent from each other. 
However, it is a matter of fact that 
a meter simple pendulum beats virtually the second 
in each oscillation, a `surprising' coincidence
that has no physical reason. 
We shortly review the
steps that led to the choice of the meter as
the ten millionth part of the quadrant of the meridian, and 
rise the suspicion that, indeed, the length of the
seconds pendulum  was in fact the starting point in establishing
the actual length of the meter.
\end{abstract}

\vspace{-12.0cm}
\begin{flushright}
         { ROMA 1\ -\ 1394/04\\
         December 2004}
\end{flushright}
\vspace{+12.0cm}

\section{Introduction}
Presently, the unit of length in the International System 
(SI)~\cite{SI}  
is a derived unit, related to the unit
of  time via the speed of light, assumed to be constant 
and of exact value $c = 299\,792\,458\,$m/s.\footnote{A meter 
is the distance covered 
by light in vacuum in  $1/299\,792\,458$ of a second.} 
The unit of time is related to the period of
a well defined electromagnetic wave.\footnote{A second is
equal to the duration of $9\,192\,631\,770$ periods of the
radiation corresponding to the transition between two hyperfine levels
(F=4, M=0 and F=3, M=0 of the fundamental status $^2S_{1/2}$) 
of the atom Cesium 133.}
The proportionality factors that relate conventional 
meter and second 
to more fundamental and reproducible quantities
were chosen
to make these units back compatible to the original units of the
Metric System. 
In that system the second was defined as 
$1/86400$ of the average solar day and the meter
was related to the  Earth meridian
(namely $1/10\,000\,000$ of the distance between the pole
and the equator on the Earth's surface). 
Therefore, meter and second were initially not formally
related to each other. 

It is then curious to 
discover --- we say `discover' because this simple observation
has surprised us, as well as many colleagues and friends --- 
that a simple pendulum 
of one meter in length  `beats' the second, in the sense that
each oscillation, i.e. half period, takes approximately one second.

A simple exercise (see Appendix A)
shows that there is no deep physical
connection between the two quantities. 
Other explanations are 
a mere, indeed remarkable 
 numerical coincidence, or 
the fact that the `first meter' was already somehow a derived unit ---
although never explicitly stated in official documents ---
in the sense that particular fraction of that particular 
geometrical parameter of Earth was chosen to approximately match the 
length of the pendulum that beats the second.

In order to understand how believable the guess 
of an initially derived meter is,
we have gone through the steps that led to the several proposals 
of a unit of length taken from nature. We have learned 
an interesting circular story of a unit of length that begins 
and ends with the second. At the beginning the unit of length
was bound to the second by gravitational acceleration, taken 
at a reference point on the Earth's surface.
Now the meter is bound to the second by the speed of light. 
In between (and somehow in parallel, at the very beginning) 
there have been the several attempts to define {\it ``a unit of  
length that does not depend on any other 
quantity''}\,\footnote{All English quotes not
referring to English bibliography are translation 
by the authors.}~\cite{Comm2}: initially
from the size of Earth; then from a conventional 
platinum bar; finally from the wavelength of a well defined
electromagnetic radiation. 

The recent relationship between the meter and the second is quite well known. 
Much less known is the fact that, 
just before the French Academy definition of the meter as a fraction of the 
Earth meridian, most scientists had considered the length 
of the pendulum that beats the second to be the most suitable
unit of length. That consensus had steadily increased during the 
18th century. However, in spring 1791 there was a metrological revolution 
inside the French Revolution.
The result was a unit of length equal to a quadrant of the 
Earth's meridian (with practical unit a convenient decimal 
sub-multiple, namely $1/10\,000\,000$). 

Our interdisciplinary work wants to reason the plausibility 
of a  meter initially bound to the second.

\section{From anthropomorphic units to 
 units taken from nature}\label{sec:physical_units}
It is universally accepted that the first important stage in the
development of metrological 
concepts related to measures of length 
is the anthropomorphic
 one, in which
the main units of measurement are the parts of the human 
body~\cite{Berriman,Kula}.
As the sociologist and historian of metrology 
Witold Kula puts it, {\it ``man measures the world with
himself''}~\cite{Kula} --- a variation of Protagoras' 
{\it ``man is the measure of all things''}.
 It is a very ancient and primitive approach. 
Certainly, even the first people who adopted such units
must have been aware that the length of their own feet 
or fingers was different from their neighbor's ones. But initially
 such personal differences did not
 seem important, given the low degree 
of accuracy required in measurements in that social context. 

Later the anthropomorphic approach reached a first level of abstraction,
characterized by {\it ``the shift from concrete
representations to abstract ones, from `my or your finger' to `finger
in general' ''}\,\footnote{The earliest 
measurement standard we have evidence of
is the {\it Egyptian cubit}, 
the length of the forearm from elbow to fingers,
realized around 2500 B.C. in a piece of marble of about 50
centimeters~\cite{Dilke}.}~\cite{Kula}.
Nevertheless, even when the stage was reached of conceiving
measurement units as abstract concepts, differences in establishing
the value of these units remained, depending on region 
or time~\cite{Hallock,Petrie,Richardson,Agnoli}.

Only in the eighteenth century,
 with the consolidation of the experimental method
on  one hand, and the drive towards international co-operation 
and trading 
 on the  other, 
 strong emphasis was placed for the first time on the need
for standardized units. In fact, the plethora of units in use
in  the various countries, and even in different towns and 
regions of the same country,
made it difficult for researchers at
different places, observing  the  same  physical phenomenon, to
 quantitatively compare their results. 
In addition, the coveted improvement in international 
commerce was in this situation
strongly hampered. 

Changing a system of weights and measures (and currency) 
in a country encounters notoriously much resistance, 
due to reasons of several natures: {\it practical}, like having to renew
all instruments of measurement and converting values between 
old and new systems in order to be back compatible with all the
information (measurements, book keeping and contracts) 
acquired in the old system; {\it sociological}, 
due to human tendency to make profit of somebody else's
ignorance or good faith 
(the recent switch from national European 
currencies to Euro is a good example of it, especially in the authors'
country); {\it psychological}, that, apart from a general inertia
to changes of a large fraction of people, is objectively
due to the fact that adults have difficulty
 in modifying their mental representations
(and, taking again the Euro introduction as an example, 
we authors still use to roughly convert 
Euro values into `thousand Lire' ---  the practical
unit of currency in Italy before the Euro --- 
multiplying the Euro value
by two). 
National laws can help to impose the new system,\footnote{With 
this respect, changes of currencies 
are easily ruled by central banks, 
especially in modern times in which banknotes and coins have no 
intrinsic value.} 
but the process of changing can take decades and 
sometimes it can simply result into a fiasco.\footnote{In history 
there are plenty of examples of
this kind, like the difficulty of France to establish  
the decimal metric system
or the failed Carolingian and Renaissance reforms~\cite{Agnoli}.
 Just to make a recent and practical example,
Italy has adopted the International System (SI) since 
the mid seventies, 
and other units were banned by law. Nevertheless, 
after thirty years, though the SI unit of power is the Watt, 
car power is quoted in HP (yes, official 
documents do have kW, but drivers, sellers and media only
speak of HP), centralized home heating power in kCal/h,  
(but small electric heaters are given in Watt) 
and air conditioning cooling power in Btu/h. 
What is bad is that, contrary to 
centimeters and inches, where people know that the units 
measure the same thing but citizens of
different countries  have a mental representation in
either unit, the average Italian does not even know that all
the above units measure the same thing 
and practically none has
a mental representation of a Btu/h,  arrived to us 
in mass with air conditioners in the last few years.
Therefore, people don't even suspect that it is possible
to convert Btu/h to Watt 
to have a better perception of what 7000 Btu/h might mean
(a cooling power of about 2\,kW).
} 

The problem becomes even more difficult to solve when one aims 
to reach an international standardization of units. 
To the above mentioned problems, we have to add
{\it national pride} that refuses to accept foreign rules,
no matter how reasonable they might be. Indeed, a similar 
regional pride existed in the eighteenth century 
also among towns and regions of the same country. 
Any attempt of standardization
would have then been perceived as an imposition of the most 
powerful town or region over the 
others.\footnote{For example, this was one of the possibilities 
envisaged in France before the metric system: to extend to all France
the system used in Paris. This was still the proposal of 
Joseph-J\'er\^ome Lalande in April 1789. 
And he strenuously defended it later
against the meter. Though his figure is often presented 
as conservative, for his opposition to the meter,
we have to admit that Lalande was quite right in proposing to
base the unit of length on a physical standard, like the 
Paris {\it toise}, rather than on the size of Earth. 
This is exactly what happened 
one century later, when in 1889, having metrologists realized that
a practical unit based on Earth 
was not accurately reproducible as required for precision measurements
and, most important, the definition itself was basically flawed, 
as we shall see at the end of this paper.
The definition of the unit of length
was then solely based upon the platinum standard, 
with no reference to Earth any more.
Perhaps what Lalande underestimated was the
psychological driving force of standards taken from nature
that, with all the problems he correctly spotted 
(see Ref.~\cite{Alder}%, p. 95
), was crucial to reach, soon or later,
some national and international agreement.}
This is the reason that led several scientists and thinkers
to seek for units that were not just as arbitrary 
as any materialization of the parts of the human body. 
Standards `taken from nature' 
({\it ``pris  dans  la  nature''}~\cite{Comm2}) 
were then seen as a possible chance to achieve 
a reform of units of measurements acceptable to all citizens of the country
and possibly to all nations.
Besides these sociological aspects, well chosen 
units based on nature have advantage of being constant with time 
and reproducible in the case that the practical standards
are corrupted or destroyed.\footnote{It is a matter of fact that
ancient standards are lost forever and the interpretation
of data taken with those units can only be guessed somehow.
A relative recent episode of an important set of
standards lost by accident is the fire in 
the British Houses of Parliament in 1834 (see e.g. 
\cite{ukmetrication}).}

The  problem of having a universal system of 
units of measurement was first tackled seriously about 
at the same time
in France, Great Britain and United States
 at the end of the 18th century, though the only
program eventually 
accomplished was that launched by the French National Assembly 
at the time  of  the French Revolution.

The French program started at the 
beginning of 1790 when the Academy was entrusted for the reform 
by the National Assembly.
As the historian Alder 
puts it~\cite{Alder}, one of the aims of 
the French scientist was to facilitate communications among scientists, 
engineers, and administrators: their {\it ``grander ambition was to 
transform France - and ultimately,
the all world - into a free market for the open exchange of 
goods and information''}. 
(For some `external' influences --- economical,
philosophical, political and cultural --- 
on the works of the French Academy see e.g. 
Refs.~\cite{Agnoli,Hahn,Rothschild}.) 
However, for the French legislator the issue had primarily
a sociological priority, as can be perceived 
in the project of the 1793 decree~\cite{luglio1793} in which {\it citoyen} 
Louis Arbogast requires the attention of the 
National Assembly on a ``{\it subject of universal beneficence}, 
 [having been] {\it the uniformity of weights and measures
for a long time one of the philanthropist wishes;
claimed at once by the sciences and the arts, by 
commerce and by the useful man who lives of the work of his hands
and who,  the most exposed to frauds, is the less capable of 
bearing their effects}.''  

The success of the French attempt to create a universal
system  of  units  of  measurement was due to several
reasons that, besides the driving illuministic spirit, 
certainly include the political drive to support
the project with education and imposition, even 
when the revolution had ended. But perhaps the main
reason that forced rulers to be so determined 
was the special chaotic situation of 
units of measurement there at that time~\cite{Zupko1}, 
where about 800 different names for measures 
have been 
estimated, whose units
varied in different towns, for a total of about 
250000 differently sized units~\cite{HistMeas}. 
{\it ``It is quite evident that 
 the diversity of weights and measures
 of different countries, and frequently in the same province, 
are a source of embarrassment in commerce, in the study of physics, 
in history, and even in politics itself; the unknown names of 
foreign measures,
 the laziness or difficulty in relating them to our own 
give rise to confusion 
in our ideas and leave us in ignorance of facts which 
could be useful to us''}, complains Charles de La Condamine~\cite{Condamine}.  

Sometimes order outsprings from chaos. That was just the case
with metrology in France.

Remarkably, the American, British and French  attempts of reform 
were originally based on the length 
of a pendulum that takes one second per swing. However,
as it is well known, the French metric system was finally based
on the size of Earth. But the decision to switch from the
pendulum to the meridian was so sudden and hurried, especially
when analyzed after two centuries, that it looks 
like a {\it coup de main}. 
Therefore, before describing the steps that led to the definition of the
meter based on the Earth meridian, let us shortly review the 
several proposals of relating units of length to the period 
of the pendulum. 

\section{The seconds pendulum}\label{sec:sp}
The idea of basing units of length on nature
 had been advocated
far before it reached definitive success with the
advent of the French Revolution. Though there were also 
proposals to relate the unit of length to the size of Earth
(see Section \ref{sec:metro}),
the unit that came out quite
{\it naturally}\,\footnote{The adjective `natural' has been
quite misused in the context of choosing the fundamental 
unit of length, calling natural what seems absolutely arbitrary to
others, as we shall see in the sequel.}
 --- or at least this was 
the proposal that had most consensus --- 
was the length of a pendulum 
oscillating with a given, well defined period. 

This is not surprising. After the first intuitions 
and pioneer studies
of Galileo Galilei at the end of the 16th  century
and the systematic experimental and theoretical researches 
of several scientists throughout
the   17th century, the properties of the pendulum
were  known rather well. The practical importance of the principle of the
pendulum was immediately recognized, and the first pendulum clock 
was realized in 1657 by  Christian Huygens. In particular, it was known
that the period of `small oscillations' of a simple pendulum
at a given place depends practically only on its length (see Appendix A). 
In other words, the pendulum was seen as an object capable to 
relate space to time~\cite{Mersenne,Huygens}.
Therefore,
discovering the possibility of grounding the unit of
length, imperfect and arbitrary since ever, to something 
regular and constant, as the alternation of days and nights,
must have been seen with enthusiasm by many scientists. 

At that time there was little doubt about what the 
unit of time should be.
The rotation of 
Earth had since ancient times provided a reference
for  units  of time, such as seconds or hours. 
The latter stem from the
subdivision  of  the day in 24 stages (12 during the daylight 
and 12 during the
night) made first by the ancient Egyptians~\cite{Hallock,Agnoli},  
and that has its roots in 
the culture of the ancient Babylonians. 
The subdivision of hours in 60 minutes of 60 seconds had become 
of common use after medieval astronomers introduced it 
in the middle of  1200, in analogy to the ancient subdivisions 
of the degree in  60 minutes of 60 seconds 
(the name second derives from the Latin {\it secundus} and refers
to the fact that the second is the `second' subdivision of 
`something', either the degree or the hour). 

It had been experimentally observed that 
one of these customary  
subdivisions of the day --- and indeed the closest to the human 
biological scale\footnote{It is not by chance that the 
smallest historical unit of time with proper name 
is approximately of the order of magnitude of the 
human heart pace.} ---
was obtained by a pendulum  
having the length in the human scale and easy to measure
(about 25 or 100 centimeters, depending on 
whether the period or half the period was considered). 
Therefore  it seemed quite
`natural' to use such a length as a unit. In particular, there was quite
unanimous agreement in associating the second
to a single swing of the pendulum, thus selecting the 
about 100 cm solution.\footnote{This choice is not surprising. 
Try to build yourself a 25 cm and a 100 cm pendula with 
a piece of string and a little weight, 
and you do not need to be an great  experimenter to 
realize that, if you want to use one of them to define 
a unit length, you would prefer to work with the longer one.}
That pendulum was called {\it seconds pendulum} 
(also {\it second pendulum}, or  {\it one-second pendulum}).

The first official proposal of basing the unit of length on the pendulum 
 was advanced by the Royal Society 
in 1660, after a suggestion  
by Huygens and  Ole R\o mer (based also on
a study of Marin Mersenne published in Paris in 1644~\cite{Mersenne}).
The proposal was followed by an analogous
suggestion by Jean Picard in 1668. A (perhaps) independent proposal 
was raised by Tito Livio Burattini in 1675, 
who called the proposed unit
`meter' and related different units in a complete system (see Appendix B).

In April 1790, one year before
the work of the commission that finally decided
to base a unit of length on the dimension of Earth,
a project based on a unit of length determined by the 
seconds pendulum at the reference latitude of $45^o$
was presented to 
the National Assembly by Charles Maurice de Talleyrand~\cite{Talleyrand},
upon a suggestion by Antoine-Nicolas Caritat de Condorcet.

Just a few months later a
{\it Plan for establishing uniformity in the Coinage,
Weights, and Measures of the United States}~\cite{Jefferson}
was presented at the other side of the Atlantic to the 
House of Representatives by USA Secretary of 
State Thomas Jefferson\footnote{Jefferson
states to have read the Talleyrand's report to the National 
Academy of France when his report was practically ready. 
Yet the {\it ``proposition made by the Bishop of Autun''}  --- this
way Talleyrand was known ---  convinced him 
to change the reference latitude of the pendulum from 
38$^o$, {\it ``medium latitude of the United States''}, 
to $45^o$~\cite{Jefferson}.}
(he became later
the third president of the United States of America). 
Again, the unit of length was based on the 
regular oscillation of a pendulum,
though the technical solution of an oscillating rod rather than a 
simple pendulum was preferred.\footnote{A homogeneous rod of length $l$
oscillating from one end behaves as a simple pendulum of length $2/3\,l$.
Therefore Jefferson's {\it second bar} was 3/2 the seconds pendulum, 
i.e. about 150 cm.} 

An analogous reform of the system of weights and measures was
discussed in the same years in the British Parliament. 
There too the  seconds pendulum was proposed, 
obviously with London latitude as reference, 
advocated by  Sir John Riggs Miller\footnote{Indeed, there 
were contacts
between Talleyrand and Miller to collaborate towards a common
solution. But due to technical and political events, 
the most relevant among them being certainly 
the French choice of the meridian,
the projects based on the pendula were put aside in France, Great Britain
and United States between 1790 and 1791, as we shall see later.}~\cite{Alder,AmChSoc}. 
The seconds pendulum was also supported by German scientists~\cite{Alder}.
\begin{table}
\begin{center}
\caption{\small Old French units~\cite{Proot}. The metric conversion 
is fixed by the French law of 10 December 1799, that established 
the meter to be equal to 
{\it 3 pied and 11.296 lignes}, i.e. 443.296 {\it lignes} 
(we give only the first six 
significant digits).}
\begin{tabular}{lcc}
\\
\hline
{\it Name} & {\it System equivalent} & {\it Metric equivalent} \\
\hline 
ligne [{\it line}] &            & 2.25583\,mm \\                 % 2.255829  0.00225582906229698
pouce [{\it inch}] & 12\,lignes & 27.0699\,mm \\                 % 27.069949 0.027069948747563
pied (de Roy) [({\it Royal}) {\it foot}] & 12\,pouces & 32.4839\,cm\\  % 32.4839385 0.324839384970764
toise [{\it fathom}] & $6\,\mbox{pieds}=864\,\mbox{lignes}$ 
       & 194.904\,cm \\                  % 194.9036310 1.94903630982459
leiue postale [{\it postal league}] & 2000\,toises & 3898.07\,m \\ % 3898.072620 3898.07261964917
\hline
\end{tabular}
\label{tab:french_units}
\end{center}
\end{table}

As a matter of fact, at the time the French Academy of Sciences
had to choose the unit of length, the seconds pendulum 
seemed the most mature candidate for the unit of length.
Moreover, with some diplomatic work concerning the choice of the reference 
parallel --- and Talleyrand was the right person for the job ---, 
there were good chances to reach an agreement 
among France, Great Britain and United States.\footnote{Jefferson 
had already accepted the French proposal of the 
45th parallel, because {\it ``middle term between the equator 
and both poles, and a term which consequently might unite the nations 
of both hemispheres, appeared to me well chosen, and so just
that I  did not hesitate a moment to prefer it
to that of $38^0$''}~\cite{Jefferson}.} 

As far as the length of the seconds pendulum is concerned,
during the 18th century
its value was known with sub-millimeter accuracy in several places
in France and 
around the world, often related to work of rather 
famous people like Isaac Newton, Mersenne, 
Giovan Battista Riccioli, Picard, Jean Richer,
Gabriel Mouton, Huygens, Jean Cassini,  Nicolas Louis de
Lacaille, Cassini de Thury
and La Condamine. 
For example, in 1740 Lacaille and Cassini de Thury had measured
the length of the seconds pendulum in Paris 
(48$^o\,50^\prime$ latitude), obtaining a value of  
440.5597 lignes (see conversion Table \ref{tab:french_units}), 
corresponding to 99.383\,cm. 
Newton himself
had estimated the length of the seconds pendulum at 
several latitudes between 30 and 45 degrees (see Ref.~\cite{Jefferson}):
his value at 45 degrees was 440.428 lignes, i.e. 99.353\,cm. 
A measurement at the equator, made by  La Condamine during the 
Peru expedition~\cite{Geodesy}, gave   
439.15 lignes (99.065\,cm).

\section{The Earth based units of length and the 
birth of the metric system}\label{sec:metro}
In the middle of 1790 there was quite an 
international convergence towards
a unit of length based on the pendulum. 
Nevertheless,
at the beginning of spring of the following year a different
unit of length was chosen by the French Academy of Sciences, 
thus signing the end of the seconds pendulum as length standard 
and causing a setback of the international cooperation
on units of measurements. 

\begin{table}
\begin{center}
\caption{\small Chronology of units of length based on nature:
from the seconds pendulum to (a fraction of) the light second.}
{\small 
\begin{tabular}{ll}
\hline
%\multicolumn{2}{c}{\bf Pioneering ideas}  \\
1644       & Mersenne writes about basing the unit of length upon
             a pendulum length. \\
1660       & The length of the {\it seconds pendulum} is proposed for unit of length\\
           &  by the Royal Society, upon a suggestion of  Huygens and  R\o mer. \\
1668       & Picard proposes the {\it universal foot}, equal to 1/3 of the length \\
           & of the  seconds  pendulum. \\
1670       & Mouton proposes a unit of length equal to {\it one minute of 
                Earth's arc}\\
           & along a meridian (equal to present {\it nautical mile}). \\
1675       & The {\it catholic} (i.e. universal) {\it  meter}, 
             equal to the length of a seconds\\
           & pendulum, is proposed by Burattini, together with a a complete \\
           & system of inter-related units based on the second. \\  
1720       & J. Cassini proposes the {\it radius of Earth} as unit of length 
             (whose \\ 
           & practical unit would be its ten millionth part). \\
1774       & La Condamine proposes a unit of length equal to the length \\
           & of the seconds pendulum at the equator (439.15 lignes). \\
1790, April & Talleyrand proposes to the National Assembly a unit of length equal \\
            & to the seconds pendulum at $45^\circ$ latitude (440.4 lignes). \\
1790, July  & Jefferson proposes to the U.S.A. House of Representative a unit of\\
            & length equal to a {\it second rod} at $45^\circ$ latitude.\\ 
\hline
1790, August & Talleyrand's proposal is recommended to king  Louis XVI  by the\\
             & Committee on Agriculture and Commerce. As a result, the Academy of \\
             & Science is entrusted of the reform of measures and weights. \\
1790, October 27 & A first commission recommends a {\it decimal scale} for all measures,\\
                 & weights and coins. \\
1791, March 19 & A second commission, set up on February 16, chooses the {\it quarter 
                 of} \\ 
               & {\it meridian}
               as natural unit of length, and its $1/10\,000\,000$ as practical unit.\\
1791, March 26 & The National Assembly accepts the recommendations of the commissions.\\
1791, April & The Academy of Sciences initially confides the new measurement of the\\
            &  meridian to M\'echain, Legendre and J.-D. Cassini
               (the latter two resigned).\\
1792, June  & Led by Delambre and M\'echain, 
              the {\it meridian expedition} finally {\it begins} \\
1793, May 29 & A third commission provides a provisional length of the 
               {\it meter} as  443.44 \\
           &   lignes (based on the 1740 measurements  by Lacaille 
             and Cassini de Thury). \\
1793, August 1 &  The provisional meter is adopted by decree.\\ 
1795, April 7 & The {\it Decimal Metric System}
                (also including including units of weight, \\
              & surface and volume --- but not the unit of time) is established by law. \\
1798,       & The meridian expedition ends.\\
1799, March & An international commission fixes the {\it length of the meter} 
              in 443.296 lignes.\\
1799, June 22 & The meter standard, a  platinum bar, is officially presented in Paris.\\ 
\hline
1812        & A hybrid system of measurements is decreed by Napoleon, in which \\
            & old names can be used for non-decimal multiples and sub-multiples\\
            & of the meter (e.g. 1 toise for `2 meter', etc.). \\
\hspace{-1.0mm}[1832 & The {\it second} is adopted in the Metric System, 
            breaking the decimal scheme.]\\  
1837        & The 1812 decree is repealed, banning old names in France. \\
1840        & Non metric system units become penal offence in France.\\
1889        & The meter is redefined as the length of the 1799 platinum bar, with \\
            &  no longer reference to the Earth meridian. \\ 
1960        & The meter is redefined as $1\,650\,763.73$ wavelengths of the \\
            &  radiation corresponding to the $2p_{10}$-$5d_{5}$ transition of 
               Krypton 86. \\
1983        & The meter is redefined as the distance traveled in vacuum  by \\
            & electromagnetic waves in $1/299\,792\,458$ of a second. \\
\hline 
\end{tabular}
}
\label{tab:milestones}
\end{center}
\end{table}
According to revolution style, the pace was very rapid 
(see central frame of Table \ref{tab:milestones}). 
In August 1790 the French National Assembly  
entrusted  the reform to the Academy of Sciences. 
The Academy nominated  a preliminary 
commission,\footnote{The Commission was made up of Jean Charles Borda, 
 Condorcet, Joseph Louis Lagrange,
Pierre Simon de Laplace and Mathieu Tillet.
}
which adopted a decimal scale\footnote{Historians generally agree 
that Mouton attempted the first metric system in 1670 when 
he proposed that all distances should be measured by
means of a decimal system of units~\cite{Mouton}.} for
all  measures, weights and 
coins.
The commission presented its report on 27 October 1790.
A second commission\footnote{This commission was made up of Borda, 
Lagrange, Laplace, Gaspar Monge and Condorcet. They worked 
in close contact with Antoine Laurent Lavoisier\,\cite{Grimaux}. 
}
was charged   with choosing
the  unit  of length. The commission was set up on 
16 February 1791 and reported to the 
Academy of Sciences on 19 March 1791.
On 26 March 1791 the National Assembly 
accepted the Academy's proposals of the decimal system and
of a quarter of the meridian
as  the  basis  for  the  new system and the adoption of the 
consequent immediate unit.

The guiding ideas of the French scientists are well expressed in the 
introduction to the document presented to the Academy:
\begin{quote}
{\small {\sl
The idea to refer all measures to a unit of length taken from 
nature has appeared to the mathematicians since they learned the 
existence of such a unit as well as the possibility to establish it: 
they realized  it was the only way to exclude any arbitrariness
from the system of measures and to be sure to preserve it unchanged 
for ever, without any event, except a revolution in the world order, 
could cast some doubts in it; they felt that such a system did not 
belong to a single nation and no country could flatter itself by 
seeing it adopted by all the others. \\
\mbox{}\hspace{0.4cm}
Actually, if a unit of measure which has already been 
in use in a country were adopted, it would be difficult 
to explain to the others the reasons for this preference that 
were able to balance that spirit of repugnance, if not 
philosophical at least very natural, that peoples always 
feel towards an imitation  looking like the admission of a 
sort of inferiority. As a consequence, there would be as 
many measures as nations.} (Ref.~\cite{Comm2}, pp. 1-2)
}
\end{quote}
Three were the 
candidates considered by the commission:
\begin{itemize}
\item
the seconds pendulum;
\item
a quarter of the meridian; 
\item
a  quarter of the equator.
\end{itemize}
The latter two units are based on the dimension of Earth. 
Indeed, Earth related units had had also quite a long history,
though they were not as popular as the seconds pendulum,
probably because their intrinsic difficulty
to be determined.\footnote{Any person can easily 
determine the length of a seconds pendulum within the percent level.
Surveying the Earth is a problem more difficult by orders of magnitude.}
Mouton had suggested in 1670~\cite{Mouton}
the unit that we still use in navigation and call now {\it nautical mile}: 
the length of one minute of the Earth's arc along a meridian, 
equal to 1852\,m.\footnote{The unit of speed consistent with it
is the  {\it knot}, corresponding to 1 nautical mile per hour. 
It is particularly suited in navigation (and hence the name).
For example, a ship that sails at 30 nodes along a meridian 
travels one degree in latitude in two hours.} 
In 1720  the astronomer Jean Cassini had proposed 
the radius of Earth~\cite{Cassini}, a `natural' unit
for a spherical object (he had also indicated the one ten-millionth part 
of the radius as the best practical unit). 
However, neither of these old, French proposals are mentioned 
in the report of the 
commission.\footnote{The Mouton's minute of meridian is just one of 
the possible subdivision of the meridian, namely the 324000-th part 
of the quarter of the meridian: we understand that the subdivision
of the right angle in 90 degrees and each degree in 60 minutes and 60 seconds
was judged `unnatural' by the acad\'emiciens 
because not decimal.  We might guess that
 the radius of Earth was not considered for the `obvious 
difficulty' to make an 
{\it immediate} measurement from the center of Earth to its surface. 
However, it should be similarly obvious that 
it was also impossible to measure 
all other quantities (diameter, meridian, equator)
in an immediate way. We shall come back to this point 
in section \ref{conclusions}.
} 

The quarter of the equator was rejected, 
mainly because considered hard to 
measure\footnote{The difficulty in measuring arcs of the equator
is not only related to perform measurements in central 
Africa or South America.
It would have required precise measurements of differences in longitude 
along the equator, and measurements of longitude are intrinsically much
more difficult than measurements of latitude, because the former
rely on  absolute synchronizations of clocks in different places, 
and that was not an easy task at that time. 
(For a novelized account of those
difficulties, see Dava Sobel's {\it Longitude}~\cite{Sobel}. 
A classical novel in which practical ways to 
measure latitude and longitude are well
described is Jules Vernes's {\it Mysterious Island}.)}
 and somehow `not democratic'.\footnote{Several of the claims
and the slogans of the {\it acad\'emiciens} show a certain
degree of na\"\i veness (frankly a bit too much
for such extraordinary clever people they were: 
the suspicion that they had 
hidden purposes in mind is almost unavoidable). 
We shall come back to this point in section \ref{conclusions}.}
\begin{quote}
{\small {\sl
So, we believe we are bound to decide to assume this kind of unit 
of measure and also to prefer the quarter of the meridian to the 
quarter of the equator. The operations that are necessary to 
 establish the latter could be carried out only in countries 
that are too far from ours and, as a consequence, we should have
 to undertake expenditures as well as to overcome difficulties 
that would be superior to the advantages that seem to be promised.
 The inspections, in case somebody would like to carry them out,
 would be more difficult to be accomplished by any nation, at least
 until the progress of the civilization reaches the peoples 
living by the equator, a time that still seems to be unfortunately
 far away. The regularity of this circle is not more assured 
than the similarity or regularity of the meridians. 
The size of the celestial arc, that corresponds to the 
space that would be measured, is less susceptible to be 
determined with precision; finally it is possible to state 
that all peoples belong to one of  Earth's meridians, 
while only a group of peoples live along the 
equator.} (Ref~\cite{Comm2}, pp. 4-5 )}
\end{quote}
The pendulum was rejected after a long discussion (two pages
over a total of eleven of the document  --- the quarter of
equator is instead ruled out in less than half a page). 
As a matter of fact, the commission acknowledges that 
\begin{quote}
{\small {\sl
the length of the pendulum has appeared in general to deserve preference; 
it has the advantage of being the easiest to be determined, 
and as a consequence to be verified.}
 (Ref~\cite{Comm2}, p. 2)
} 
\end{quote}
Then the report specifies that the pendulum 
should be a simple pendulum at the reference latitude 
of $45^o$, because {\it ``the law followed by the lengths
of simple pendula oscillating with the same time 
between the equator and the poles is such 
that the length of the pendulum at the forty-fifth parallel is 
precisely the 
mean values of all these lengths''}.\footnote{Actually, 
such a length is not
`precisely' the average over the lengths at all parallels, but 
only a very good approximation. In fact, the net 
gravitation acceleration $g$ does not vary linearly with the
latitude, but follows the following law~\cite{Geodesy}:
$$g/(\mbox{m/s}^2) = 9.7803185(1+0.005258895\,\sin^2\phi-
0.000023462\,\sin^4\phi)\,, $$
where $\phi$ is the latitude (the theoretical formula
is valid at the ellipsoid surface and, as is customary in geodesy,
$g$ is the sum of the effects of 
gravitation and centrifugal forces).
}~(Ref.~\cite{Comm2}, p. 3)

 There was still the problem of 
the reference time of the pendulum, since
the second was considered 
an  ``{\it arbitrary subdivision of this natural unit} 
[the day]''~\cite{Comm2}. 
But a possible way out was envisaged:
\begin{quote}
{\small {\sl
In reality, we could avoid the last inconvenience
 taking as unit the
hypothetical pendulum that made just one oscillation in one day;
its length, divided by ten billion parts, would give 
a practical unit of measurement of about twenty seven pouces} 
[27\,pouces\,$\approx\,$73\,cm]{\sl; and 
this unit would correspond to the pendulum that makes 
one hundred thousand oscillations in one 
day.}\footnote{Ten billion is the square of one hundred thousand 
(the length of the pendulum is proportional 
to the square of its period). 
Seen with modern eyes, it looks a bit bizarre 
that this `hypothetical pendulum', sized almost
twenty  times the distance Earth-Moon,
would have been natural, while the second wouldn't. 
(By the way, for those who like to understand all digits:
73\,cm comes from rounding to the pouce a
length that, directly rounded to the centimeter, 
would be 74\,cm.)
} 
(Ref.~\cite{Comm2}, p. 4)
} 
\end{quote}
Essentially, the report of the commission does not 
provide any specific weakness of the 
pendulum\footnote{It should be noted that a unit of length based
on the pendulum has intrinsic problems, like the dependence 
of its period on temperature, latitude and above sea level, 
plus other more technical issues, considered also in Jefferson's
document~\cite{Jefferson}. But, in the part of the report
in which the seconds pendulum is discussed and rejected 
as unit of length, the French commission
does not seem concerned at all with this kind of 
physical questions. Only later (p. 9), when they propose the 
pendulum as ancillary reference of length, they specify that 
the pendulum should beat {\it ``at the sea level, on vacuum and
at the temperature of melting ice.'}'~\cite{Comm2}
} and, finally, the choice of the quarter 
of the meridian is justified only in terms of `naturalness',
 as it was perceived by Borda and colleagues:
\begin{quote}
{\small {\sl
\mbox{[\ldots]} one would still have to include an 
heterogeneous element, time, or what is here the same thing,
the intensity of the gravitational force at the Earth's surface.
Now, if it is possible to have a unit of length that does not depend 
on any other quantity, it seems natural to prefer it.\footnote{Our 
note: to be precise, that is not right ``{\it the same thing}.''
The pendulum relates space and time via the net gravitational
acceleration $g$. Therefore the included heterogeneous elements 
are two: time and acceleration. A similar situation happens now, 
where the meter is related to the second via the speed of light.} \\
\mbox{[\ldots]}\\
Actually, it is much more natural to refer the distance between 
two places to a quarter of one of the terrestrial circles  
than to refer it
 to the length of the pendulum.\\
\mbox{[\ldots]}\\
The quarter of the Earth meridian would become then the
real unit of length; and the ten million-th part of this 
length would be its practical unit.} (Ref~\cite{Comm2}, pp. 4-5)
}
\end{quote}
It must be stressed that, however, there was still
a strong resistance from those 
scientists who preferred the pendulum~\cite{Alder,Zupko}.

The new unit was officially called `meter'\footnote{The
name `meter' comes from Greek {\it metron}, meaning 
`measure'. The first who proposed the name 
meter in the 
context of the French Academy work
is acknowledged to be 
 the mathematician Leblond in 1790~\cite{Leblond, Leblond1};
still, some historians (see e.g. Ref.~\cite{Zupko})
maintain the idea has to be originally attributed to Borda.
However, the name meter for a unit of length (that practically
coincides with the French Academy meter) was proposed  
more than one century earlier by Burattini 
(see Appendix B).} 
only two years later (see section \ref{ss:provitional}), 
in the occasion of a  report of 
a new commission, in which also for the first time an estimate
of its length was made public
by Borda, Lagrange and Monge~\cite{BLM,luglio1793}.

\section{Establishing the length of the meter}\label{lunghezza_metro}
%As just mentioned, figures for the length of the meter were officially 
%for the first time in 1793. 
Reading the {\it Rapport sur le choix d'une unit\'e de mesure}~\cite{Comm2}
we have been surprised not to find 
the expected value for the new unit of length.
 All looks as its
length was unknown and it had to be determined by the 
campaign of measurements outlined in the document.
The seconds pendulum has the same omission. But we 
imagine that the members of the National Assembly,
to which the document had to be finally read, 
were curious to know the rough
length of the unit they were going to decree. 
Instead, the commission provides only an 
estimate of the size of the unit
that would result from the ideal, 7.4 million kilometer 
long pendulum that beats the day. 
Therefore, it seems reasonable to believe  that
the length of the seconds pendulum and of the 
fourth-millionth part of the meridian were 
already known rather well, and that there was no need to specify 
their value. This was our first guess. Actually, 
 the question looks a bit more subtle at a closer look: 
though those values
were well known to scientists, 
the  acad\'emiciens kept them `secret' or, at least, they
were reluctant to provide an official best estimate of the
new unit of length to the politicians~\cite{Alder}.  
It seems to us that the reason of this reserve is closely 
related to the preference of the meridian over the pendulum.
But let us proceed with order. 

The seconds pendulum has been reviewed in section~\ref{sec:sp}.
 Comparing 
Cassini's and Newton's value, we can safely take an approximated 
value of the seconds pendulum at $45^o$ of 440.4 lignes (99.35\,cm). 
Let us now see how well the `meter' was known at the time
of the March 1791 report~\cite{Comm2}. We shall
then go through  the recommendations of 
the commission for a more accurate determination of the unit of length
and through the resulting meridian expedition.

\subsection{Measurements of the Earth meridian before 1791}
Measuring the size of Earth had been a challenging problem
for ages, since it was first realized that
Earth is spherical, i.e. 
at least by the sixth century B.C.~\cite{Geodesy}.
The most famous ancient estimate is that due to Eratosthenes 
(276--195\,B.C.), who reported a value of 250\,000 stadia, 
i.e. about $\approx 40\,000$ kilometers,
if we take 159\,m per stadium.\footnote{Due to uncertainties in
the conversion factor stadium-meter, approximations and errors
in evaluating distances and differences in latitude, the 
usually quoted value of 40\,000 kilometers obtained by Eratosthenes 
has to be considered fortuitous, being the uncertainty
on that number of the order of 10\%~\cite{Geodesy}. Anyway, 
not bad for that time (in frontier physics a completely new
measurement that provides a result with 10\% uncertainty is
considered a good achievement).}  

\begin{table}[t]
\caption{\small Some milestones in measuring the Earth meridian. $l_m$ 
stands for the length of the meter calculated as the 40\,000\,000th
part of the meridian (for some important cases, $l_m$ is
also given in lignes --- see 
Table \ref{tab:french_units} for conversion).
In the results expressed in 
the form `$\mbox{xxxxx}\times 360^o$' xxxxx stands for the
length of one degree meridian arc ($s/\alpha$ in the text).
Ancient estimates have to be taken 
with large uncertainty (see e.g. Ref.\cite{Geodesy}).}
\begin{center}
{\small
\begin{tabular}{llrcll}
\hline
Author(s) & Year & Value & Unit & km value & $l_m$ (m) \\
         &      &       &      &   & \{lignes\}  \\
\hline
 Eratosthenes          &  (III B.C.) & 250000 & stadium$^a$ 
                   & $\approx 40000$ & $\approx 1.0$    \\
     &      &       &      &    \\
Caliph Al-Mamun & 820  & 56~$^2$/{\tiny 3} & Arab mile$^b$  & 39986 & 0.9997 \\
     &      &       &      &    \\
Fernel    & 1525   &  $56746\times 360^o$ & toise &39816  & 0.9954 \\
Snellius    & 1617   &  $55100\times 360^o$ & toise & 38661 & 0.9665  \\
Norwood    & 1635   &  $57300\times 360^o$ & toise &40204 & 1.0051  \\
Picard    & 1670   &  $57060\times 360^o$ & toise &40036  & 1.0009 \\
J. Cassini & 1718   &  $57097\times 360^o$ & toise &40062 & 1.0016  \\
  Lacaille and  & 1740  & $57027\times 360^o$ & toise & 40013 & 1.00033  \\
  Cassini de Thury     &      &       &    &  & \{443.44\}\\
     &      &       &      &     \\
\mbox{[}{\it \,Lapland expedition} & {\it 1736} &{\it 57438}$\,\times\,${\it 360$^o$} 
                    & toise & {\it 40302} & {\it 1.0075}\,]\,$^c$ \\
\mbox{[}{\it \,Peru expedition}  & {\it 1745} &{\it 56748}$\,\times\,${\it 360$^o$} 
                    & toise & {\it 39817} & {\it 0.9954}\,]\,$^c$ \\
     &      &       &      &     \\
 {\bf Delambre and} & {\bf 1799}  & {\bf 20522960} & {\bf toise} 
& {\bf 40000} & {\bf 1} \\
  {\bf M\'echain}   &      &       &    &  & \{{\bf 443.296}\}\\
                         &   &[{\it \,57019}$\,\times\,${\it 360$^o$} & toise &
           {\it 40008} & {\it 1.00019}\ \  ]\,$^d$ \\
   &      &       &    &  & \{{\it 443.38}\}\\
     &      &       &      &     \\
 present value       &      & 40009152  & m &  40009.152 & 1.000229  \\
   &      &       &    &  & \{{\it 443.3975}\}\\
\hline  
\multicolumn{6}{l}{$a)$ Stadium estimated to be 159 m.}\\
\multicolumn{6}{l}{$b)$ Arab mile estimated to be 1960 m~\cite{Geodesy}.}\\
\multicolumn{6}{l}{$c)$ Values obtained at extreme latitudes, very
sensitive to Earth ellipticity.} \\
\multicolumn{6}{l}{$d)$ The entries of this line assume
 a spherical model for Earth, as for older estimates.} \\  
\multicolumn{6}{l}{\ \ \ \ The value of 57019 toises per degree is obtained dividing
the $551\,584.74$ toises of the } \\
\multicolumn{6}{l}{\ \ \ \ meridian  arc Dunkerque-Barcelona by their  
difference in latitude,
$9^o\,40^\prime\,25.40^{\prime\prime}$\,\cite{Guedj1} .}
\end{tabular}
}
\end{center}
\label{tab:meridiano}
\end{table}

The principles of measurement of the Earth parameters,
exposed at an introductory level, 
as well as  milestones of  these achievements, can be found
in Ref.~\cite{Geodesy}. The basic idea is rather
simple: if one is able to measure, or estimate somehow, 
the length of an arc of meridian ($s$) 
and its angular opening ($\alpha$), 
the length of the meridian can be determined 
as $360^o\times\,s/\alpha$,
if a circular
shape for the meridian is assumed (i.e. for a spherical Earth).  
The angle $\alpha$ 
can be determined from astronomical observations.
The measurement of $s$ is bound to the technology of the epoch
and varies from counting the number of steps in the early days 
to modern triangulation techniques~\cite{Geodesy}.
Indeed, the rather accurate measurements between the 
16th and 18th centuries 
provided the results in terms of $s/\alpha$, expressed e.g. in toises/degree
(see Table \ref{tab:meridiano}).

As we can see in Table \ref{tab:meridiano}, not exhaustive
of all the efforts to pin down the Earth dimensions, 
there was quite
a convergence on the length of the unitary meridian arc in Europe,
as well as a general consistency with older measurements. 
For example, taking the Lacaille and Cassini de Thury result, 
based on the about 950 km arc of the Paris 
meridian\footnote{The Paris meridian, now  
$2^0 20^\prime 14^{\prime\prime}$\,East,
had been the oldest zero longitude, until in 1884 
it was replaced by the Greenwich meridian, even though France and Ireland
adopted the new zero only in 1911.}
 across all France,
it is possible to calculate a value of 443.44 lignes for the
new unit of measurements, assuming a spherical Earth
($57027\times 90^o / 10\,000\,000 \times 864=443.44$). Even a conservative
estimate would give a value of 443.4 lignes, with an uncertainty
on the last digit --- a difference of three lignes (about 6 mm)
 with respect to the length of the second pendulum. That was
definitely known to Borda and colleagues. 

However, apart from experimental errors in the determination 
of the 57027 toise/de\-gree, the value of  443.44 
was still affected by uncertainties
due to the shape of Earth. At that time the scientists were
rather confident on an elliptical shape of the meridians, 
resulting from the Earth flattening at the poles. In fact, 
centrifugal acceleration due to rotation  is responsible for the bulge 
of the Earth at the equator. The resulting Earth shape is
 such that the total force (gravitational plus centrifugal)
acting on a body at the Earth surface
 is  always orthogonal to the `average' surface of the Earth. 
If that were not the case, there would be tangential forces
that tended to push floating masses towards the equator, 
as eloquently stated by Newton: {\sl 
``\ldots if our Earth were not a little higher around the equator
than at the poles, the seas would subside at the poles and,
by ascending in the region of the equator, would flood everything there.''
} (Cited in Ref.~\cite{BulgingEarth}.) 
Newton had estimated an Earth ellipticity of 1/229\,\cite{Greenberg}.

Several measurements had been done during the 18th century 
to determine the value of Earth flattening. 
In particular, there had been an enormous
effort of the French Academy of Sciences, that supported 
measurements in France as well as expeditions at extreme 
latitudes, up to the arctic 
circle and down to the equator.\footnote{The Lapland 
expedition measured 
an arc of $57^{\prime}$ crossing the north polar circle in northen Finland,
at an average latitude of $66^o\,19^{\prime}$\,N. 
The Peru expedition measured an arc of $3^0\,7^{\prime}$
at an average latitude of  $1^0\,31^{\prime}$\,S
(see \cite{stamps}
for a nice web site dedicated to the expeditions).}

The latter measurements were essential in order 
to gain sensitivity on the flattening effect. In fact, the 
unitary arc length $s/\alpha$ gives the local curvature 
along the meridian around the region of the measurements. 
As a consequence, $\rho=s/\alpha \times 360^o/2\pi$
is equal to the radius of the circle that approximates locally
the meridian ellipse. 
\begin{figure}
\begin{center}
\epsfig{file=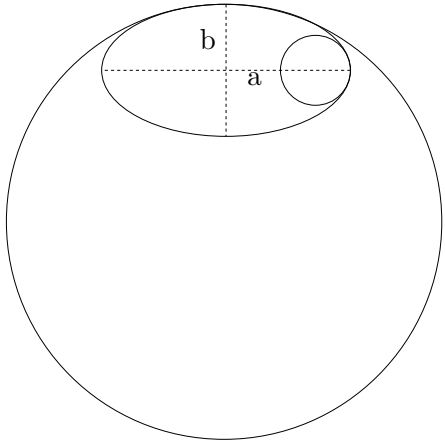,width=0.4\linewidth,clip=} 
\end{center}
\vspace{-6.1cm}
\hspace{+6.8cm}
b\\
\mbox{}
\vspace{+1.2cm}
\hspace{+7.3cm}\mbox{}\vspace{+0.2cm}
a
\vspace{+3.6cm}
\caption{\small An exaggerated representation of the ellipsoidal
Earth shape, showing local circles at the equator and 
at the pole (the ellipse is characterized by 
the semi-axes $a$ and $b$).
The ellipse of the figure has a flattening of about 1/2, 
i.e. an eccentricity of 0.87. 
(A flattening of 1/298, corresponding to the 
Earth one, i.e. a minor axis being 0.3\% smaller
than the major axis, is imperceptible to the human eye.) 
Note how the equatorial local 
circle underestimates the ellipse circumference,
while the polar one overestimates it.}
\label{fig:ellisse_cerchi}
\end{figure}
As it can be easily understood from figure \ref{fig:ellisse_cerchi},
the curvature decreases with the latitude: the radius of 
the `local circle' is minimum at the equator and maximum at the pole.
The measurements of arcs of meridian at several
latitudes (two distant latitudes are in principle
sufficient) can yield the ellipse parameters.
Comparing the result of  Lacaille-Cassini from Table~\ref{tab:meridiano}
with the results of the Lapland and the Peru expeditions 
from the same table, we see that $s/\alpha$ is indeed increasing with 
the latitude (note these measurements were quite accurate ---
 for a very interesting account of the Peru expedition
see Ref.~\cite{Smith}). The combination of these and other 
measurements gave values of the Earth flattening in the
range 1/280--1/310~\cite{Smith}, with a best estimate around
1/300, very close to the present value of 1/298
(see Table \ref{tab:dati_terra}). 
\begin{table}[t]
\caption{\small Earth data~\cite{TOE17}. The geometrical data
refer to the WGS84 ellipsoid~\cite{Smith}. Note that the generic `radius of Earth' 
$R$ refers usually to the equatorial radius, but sometimes also
to the `average' equivolume radius.  
In literature the name 
`ellipticity' is often associated to what is called `geometric
flattening' in this table, and even to the difference of 
equatorial and polar radii divided 
by their average, i.e. $(a-b)/((a+b)/2)$.
Anyway, the three different `ellipticities' give with good approximation
the same number, about $1/298$, because of the little deviation 
of our planet from 
a perfect sphere. 
 Note also that sometimes the flattening
is even confused with the ellipse eccentricity, that differs quite
a lot from flattening and ellipticity.}
\begin{center}
\begin{tabular}{ll}
\hline
equatorial radius, $a$   & 6\,378\,137\,m \\
Polar radius, $b$        & 6\,356\,752\,m \\
Equivolume sphere radius & 6\,371\,000\,m \\
Geometric flattening, $f=(a-b)/a$    & 1/298.26 \\
Ellipticity, $(a^2-b^2)/(a^2+b^2)$ & 1/297.75  \\
Eccentricity, $e=\sqrt{1-b^2/a^2}$ & 0.08182 = 1/12.2 \\
(for $f\ll 1$, $e$ is about $\sqrt{2\,f}$) & \\
% ($e^2$ & 0.006694 = 1/149.37) \\
Mass, $M$  & $5.97369\times 10^{24}\,$kg \\
Mean density & $5.5148\times 10^3\,$kg\,m$^{-3}$ \\
Normal gravity at equator & $9.7803267\,$m\,s$^{-2}$ \\
Normal gravity at poles   &  $9.832186\,$m\,s$^{-2}$ \\
$G\,M$ \ (where $G$ is the gravitational constant)& 
$3.986005\times 10^{14}\,$m$^3\,$s$^{-2}$ \\
\hline
\end{tabular}
\end{center}
\label{tab:dati_terra}
\end{table}
However, given such a tiny value of the flattening (imagine a soccer ball
squeezed by 0.7 mm), its effect on the circumference of
the ellipse is very small, of the order of a few parts 
in $10\,000$.  

To summarize this subsection, we can safely state that
the length of the  meridian, and hence
of any of its subdivisions, was known with a relatively high accuracy
decades before the 
report that recommended the unit of length equal
to $1/10\,000\,000$ part of the quarter of meridian was produced.
In particular, for what this paper is concerned, 
it was well known that the new standard
was equal to the length of the one seconds pendulum within 
about half percent.

\subsection{The 1793 provisional meter}\label{ss:provitional}
The first public figures for the new standard, together 
with the name {\it meter}, were provided by the acad\'emiciens 
in a report by Borda, Lagrange and Monge in spring 1793\,\cite{BLM}.
A project of decree for a general system of weights and measures,
also containing the cited report, was presented to the National 
Assembly in July of the same year\,\cite{luglio1793}. 
The length of the meter was obtained from the Lacaille-Cassini
measurements, with the simple calculation shown 
in the previous subsection.
\begin{quote}
{\sl \small
Its approximated value is 3 pieds 11 lignes 44/100 
present Paris measure {\rm [443.44 lignes]},
and this approximation is such that its error does not exceed
one tenth of 
\newpage
\mbox{}
\vspace{0.5cm}\mbox{}\\
\noindent
ligne {\rm [0.2 mm]}, that is sufficient 
for the ordinary use in the society. This unit will
take the name {\it meter}. 
} {\small (\cite{luglio1793}, p. 5)} 
\end{quote}
%\newpage
%\mbox{}
%\vspace{0.5cm}\mbox{}\\
%\noindent
The so called {\it provisional meter}, 
with all the resulting units of volume, 
weights and surfaces~\cite{luglio1793,BLM},
was adopted by decree the 1st August 1793.

\subsection{The 1792-1798 meridian mission}
The {\it Rapport sur le choix d'une unit\'e de mesure}~\cite{Comm2}
outlined a plan to realize the new 
system of measures. One of the items of the plan was to re-measure
the Lacaille-Cassini arc of the Paris meridian 
from Dunkerque
to Perpignan,\footnote{More precisely, the 
lowest latitude was about Collioure, 
a small town close to the 
Spanish  border.}
also extending the 
measurements down to Barcelona (a total of 
almost ten degrees in longitude,
for a length of 1075 km ---
see the map in Ref.\,\cite{Meridiano} 
to get an idea of ``{\it cette belle entreprise}''\cite{luglio1793}). 

The motivation of the Academy in support of the enterprise
was to improve the knowledge of the meridian length. 
The confidence on a possible improvement relied  
on several factors: 
\begin{quote}
{\small {\sl
the ability of the astronomers
presently involved in this job, 
the perfection that the
mathematical tools and instruments have acquired in the
last times, the magnitude of the measured circle, 
that is extended by more than nine degrees and one half, 
the advantage of being this {\rm [the meridian arc]} 
cut in the middle 
by the forty fifth parallel, all that guarantees us 
the exact and perfect execution of this
beautiful enterprise, the greatest of this kind.
(Ref.~\cite{luglio1793}}, pp. 3-4)
}
\end{quote}
The  campaign of measurements started immediately, 
apart from delays of practical nature, 
after the Academy's proposal was accepted. 
A quote is in order to report the driving spirit of the 
commission, that also shows the yearning for
 an `egalit\'e of measures',
one of the  political and 
ideological requirements of the new r\'egime 
that can't be set aside  any longer.
\begin{quote}
{\small {\sl
There is no need, in our judgment, to await the concurrence 
of other nations either to choose a unit of measurement or to begin 
our operations. Indeed,  we have eliminated all arbitrariness from 
this determination and rely only on information equally accessible 
to all nations.} (Ref.~\cite{Comm2}, p. 11)
}
\end{quote}
Two astronomers were nominated responsible of the mission:  
Jean Baptiste Joseph Delambre, in charge of the northern 
part of the arc, up to  Dunkerque, a French town 
on the North Sea, close to the Belgian border;
Pierre Fran\c cois Andr\'e M\'echain, in charge of the 
southern part. 

In principle, the task was simpler than that accomplished by
Lacaille and Cassini fifty years earlier, because 
it seemed initially possible to use much of their
work (like the triangulation stations). However, things 
were much more difficult, complicated by Revolution
and wars (see Refs.~\cite{Guedj,Alder} for a dramatic 
account of the enterprise). It was almost a miracle
that Delambre and M\'echain could meet again in Paris 
 in November 1798 alive and 
with the logbooks of their measurements.

The unexpected long duration of the mission 
 was the reason the acad\'emiciens were urged
to provide the provisional length of the meter in 1793.
In fact, ``{\it the interests of the Republic 
and of the commerce, the operations initiated on the money
and on the cadastre of France, require that the adoption
of the new system of weight and measures is not
delayed any longer}''\,\cite{luglio1793}.
 The {\it Decimal Metric System} was later
established by law on April 7, 1795,
well before the meridian mission was accomplished.

After the end of the meridian mission an international commission 
was convened to review the Delambre-M\'echain
data and to establish the length of the meter. 
In March 1799 the meter was determined in 443.296 lignes, 
also taking into account Earth flattening.\footnote{The 
Dunkerque-Barcelona arc is sufficiently large to allow an
estimate of the meridian curvatures in several sub-arcs and to make an
independent estimate of Earth flattening. That came out to be 
about one half of that based on many data sets
 from equator to Lapland. The flattening based on the
latter information was finally preferred, 
As Alder puts it ``{\it \ldots 1/150 offered the best description
of the arc as it passed through France, but they knew that the 
older data offered a more plausible picture of the overall
 curve of the Earth. They could choose consistency or plausibility.
And after some heated discussion, they chose plausibility 
and the old data.}'' Actually, the discrepancy between the 
values of flattening in different  sub-arcs
was a first indication that Earth has 
a more complicate shape than just a rotational ellipsoid,
giving rise to the concept of {\it Geoid} (see e.g. \cite{Geodesy}).}
The new standard differed by 0.114 lignes (0.32 mm)
from the provisional unit. Compared with our present 
value (see Table \ref{tab:meridiano}) one can see that the 
new result slightly worsened the knowledge 
of the meridian.\footnote{
We would like to point out that the two results would come
to a better agreement if they were treated in the same way.
In fact, correction for flattening was not applied to the 
 Lacaille-Cassini result. 
In Table  \ref{tab:meridiano} we have done the exercise of 
calculating the length of the meridian from $s/\alpha$ of 
Delambre-M\'echain, equal to 57019 toises/degree. We obtain
a resulting meter of 
443.379 lines. Since the Lacaille-Cassini arc was roughly 
similar to the Delambre-M\'echain one, we can use the ratio
443.296/443.379 as an
approximate correction factor to take into account 
Earth flattening in the  Lacaille-Cassini data. 
After the correction the meridian length becomes
40006 km and the corrected provisional meter would be 443.36 lignes
(and $l_m=1.00014\,$m), i.e. a difference of 
only 0.064 lignes (1.4 mm) 
with respect to the 1799 final meter. 
Anyway, though the two results get closer, the one 
based on the Lacaille-Cassini measurements gets also
slightly closer to the present value of the meridian length.}

The manufacture of the definitive model,
based on the results on those measurements, was completed in
June  1799.
On 22 June, the prototype of the meter was solemnly
presented
to  the Council of Elders and of the Five Hundred. 

\section{Interplay of pendulum and the meridian based standard}
\label{sec:aside}
As we have remarked several times,  officially
it  seems that the meter was invented `out of nothing', 
apparently only stemming from the slogan 
``{\it ten million meters
from  the  pole to the equator}''~\cite{Delambre2}.
A pendulum had been considered as a possible candidate
as unit of length, but it had been rejected to avoid a unit of length
depending on a unit of time. 
However, in listing the necessary operations 
in order to realize the reform of measures,
a {\it unitary pendulum}
is taken into account, as a kind of secondary standard
to reproduce the meter: 
\begin{quote}
{\small {\sl
The operations that are necessary to carry out this work are the following:
\mbox{\ldots} \\ 4th \ \ 
To make some observations in latitude forty-fifth degree to verify
 the number of oscillations that a simple pendulum, which corresponds to 
the ten millionth part of the arc of a meridian,  would accomplish 
within a day, in the vacuum, at sea level, at freezing point, so that,
 after having learned that number, this measure could be found again 
through the observations of the pendulum. In this way the advantages 
of the system that we have chosen are joined to the advantages 
that would be obtained by taking the length of the pendulum as a unit. 
It is possible to accomplish these observations before learning 
this ten millionth part. Actually, if the number of the oscillations 
of a pendulum having a determined length are known, it will be 
sufficient to learn the relation between this length and the 
ten millionth part to infer undoubtedly the investigated 
number.\footnote{The task specified in the
 4th  point was later 
committed to Borda and  Charles Augustin de Coulomb.}
} (Ref.~\cite{Comm2}, p. 9)
}\end{quote}
Therefore, while in the first part of the report the one-second pendulum
was ruled out, a one-meter pendulum\footnote{This name is not
appropriate for the pendulum mentioned in 
the 1791 document, as the name 'meter' had still to be
made official. Nevertheless, let us call it so hereafter.} appears 
in the second part. 

The fact that 
the meter pendulum had to be considered an
important tool to carry out the entire project 
can be easily inferred from the response of the President
of the National Assembly to a memory of Borda~\cite{Borda} on the 
metric system of weights and measures in 1792:
\begin{quote}
{\small {\sl You took your theory from the nature: 
among all the determined lengths you chose the two unique
 lengths whose combined result was the most absolute,
the measure of the pendulum
 and above all the measure of the meridian; and by so relating
 the first to the second with as much zeal as sagacity,
 the double comparison of time and Earth through
 a mutual confirmation, you will have the honor to
 have discovered this permanent unit for all the world,
 this beneficent truth that is going to become a new 
advantage to the nations and one of the most useful
 conquest of the equality.
} (Ref.~\cite{Borda}, p. 9)
}\end{quote}
And, again, in the July 1793 proposal of decree, it is stated that
\begin{quote}
{\small {\sl 
the academy has judged that its works were quite
well advanced, and that the arc of meridian, as well as the length
of the seconds pendulum, the weights of a cubic pied of
distilled water, were known at this moment, both from 
previous observations as well as from those on which the 
members of the commission have been working, with 
the accuracy sufficient to the ordinary usages of the society
and the commerce
} (Ref.~\cite{luglio1793}, p. 4)
}
\end{quote}
Therefore, while in the first quote of this section 
(Ref.~\cite{Comm2}, p. 9) the commission generically 
speaks of a ``{\it a pendulum of a determined length}'', it is
here clear that the referred pendulum was the seconds pendulum,
and not a pendulum based on a previous unit of length, 
like a one-toise pendulum. 
We see clearly the advantage of that choice. 
Since the seconds pendulum was quite well known, 
including the dependence of its period with latitude, 
once the oscillation time of the meter $t_m$ 
was known, 
the unitary length would have been easily determined as 
%$u_m = l_{sp}\,t_m^2/\mbox{s}^2$, 
$u_m = l_{sp}\,t_m^2$, 
where $l_{sp}$ indicates the length of the seconds pendulum
and $t_m$ is expressed in seconds. 
However, it was never stated that the seconds pendulum
and one-meter pendulum are very close in length, though
their closeness makes easier the intercalibration procedure.
%[Since $t_m$ is close to one second, if we call
%$\epsilon$ its small difference with respect to 
%the exact second, i.e. $\epsilon = t_m - 1$, 
%we obtain the approximated
%relation $u_m \approx l_{sp}\times(1+2\,\epsilon)$.]

\section{Conclusions and discussion}\label{conclusions}
The initial aim of this paper was to rise the question
that gives the title, why the semi-period of 
a  meter  simple pendulum  
is approximately equal to one second.
 In fact, it seems to us that this question
has never been raised in literature, not even at a level of a curiosity
(for some references on the subject, see e.g. 
\cite{Zupko,Guedj1,Guedj,Alder,Roche}). 

We were fully aware that mere coincidences are possible 
and that speculations based on them can easily drift to the non-scientific
domain of numerology, especially in the field of units of measures,
where the large number of units through the ages and around the world
cover almost with continuity the range of lengths in the 
human scale.
Therefore, we were searching if there were reasonable explanations
for the close coincidence pointed out in our question. 

A physical reason is  ruled out. A unit of length
defined as a fraction of a planet meridian makes the 
period of a unitary pendulum only depending on the planet density. 
But this period has no connection with the planet rotation 
period, and then with its 86400th part. 

The suspicion remains --- definitely strengthened   --- 
that, among the possible choices 
of Earth related units, the choice favored the one that 
approximated a pendulum that beats the second.

\subsection{About the choice of the meridian}
As we have remarked in Section \ref{sec:metro}, 
the {\it Rapport sur le choix d'une unit\'e de mesure}\,\cite{Comm2}
shows a certain degree of na\"\i veness. For example, 
let us take the preference
of the meridian over the equator, justified in terms of 
ease of measurement and of universality with respect to all nations of Earth. 

Let us start from the latter point, that we called `democratic'. 
The sentence  ``{\it it is possible to state
that every people belongs to one of the Earth's meridian, while
only a group of people live along the equator}'' is not more
than a slogan. It is  self-evident that every point on the Earth 
surface belongs to a meridian. A different question 
is to measure it in order to reproduce the meter with 
the accuracy required to exchange the results of precise
measurements in different places of the globe.
Even assuming that every people had the 
proper technology to perform the measurements, the country should be
extended enough  along the longitude in order the required 
measurements to be performed.
 Moreover, 
since Earth is flattened at the poles, 
at least two measurements of arc of meridian at two distant latitudes 
are needed, in order to infer the Earth ellipticity. 
Therefore, also the sentence ``{\it The operations that are necessary to 
 establish the latter could be carried out only in countries 
that are too far from ours}'', referred to the equator, 
applies also to the meridian, with even a complication for the latter:
while the determination of the equator requires only one campaign of 
triangulation, because (apart from small mass dishomogeneity) 
the circularity of the equator 
comes from symmetry arguments,
determining the meridian requires
necessarily, as it had already been done in the middle
of the 18th century, several campaigns at different latitudes.

Frankly, we find that  na\"\i veness acquires an
unintentional humorous vein
at page 8 of the document \cite{Comm2}, when, after having claimed
a few pages earlier that the choice of the meridian is `democratic',
the Paris meridian going from Dunkerque to Barcelona 
is presented as almost unique to perform
the proposed measurement:\footnote{That could be the reason
why the world best selling encyclopedia erroneously reports 
that the meter 
``{\it was originally defined as one ten-millionth of the 
distance from the equator to the North Pole 
on a line \underline{running through Paris}.}''\cite{EncartaMsn}
This could be just a minor flaw due to superficiality,
but it could also be a heritage of the anglo-saxon  
reaction to what was perceived as a French imposition.}
\begin{quote}
{\small {\sl 
One cannot find neither in Europe nor 
in any other part of the world, unless to measure 
a much wider angle, a portion of meridian that satisfies
at the same time the condition to have the extreme points at sea level,
and that of crossing the forty-fifth parallel, if one does not
take the line that we propose, or as well another more western meridian 
from the French coast, until the Spanish one. 
 (Ref.~\cite{Comm2}, p. 8)
}
}
\end{quote}
(Their precise Swiss neighbors would have no chance 
to reproduce the meter standard in their country!)
As a matter of fact, the choice of the meridian
 appeared to other countries, especially Great Britain and USA, 
as a imposition of the Paris meridian. All previous attempts
of cooperation towards an international standard of length
were frustrated, and still now we suffer of
communication problems.\footnote{As a side remark, we would like 
to point out that even the very interesting Alder's book 
{\it The Measure of  All Things}~\cite{Alder}, that has been very useful 
to us in this research, is a proof the dreams of
the acad\'emiciens are still far from coming true.  In fact 
most lengths are given in feet and inches, 
used by to American and British readers, 
but hard for the others, especially when, in translation, unavoidable
mistakes happen, as the 25 feet of page 188, that becomes 672 meters
at page 289 of the Italian edition. Nothing compared to the 
Mars Climate Orbiter disaster, but this is symptomatic 
of the troubles that disuniformity of units of measures still causes,
made worse by globalization. 
}

Let us  come  now to the concept of `naturalness',
about which we have already expressed 
some {\it caveat} above. 
Once chosen Earth as reference object upon which 
the scale of lengths has to be based, 
which of its parameters is the most natural?
For simplicity, let us consider a sphere. To a mathematician or
a physicist the natural parameter of the sphere
is the radius (that was 
basically the reason of Cassini's proposal 
mentioned in section \ref{sec:metro}). However, for a engineer
the natural parameter is the diameter, because that is what
he directly measures with a gauge in the workshop. 
The diameter is also the convenient parameter for a sphere 
seen from very far (as it could be a planet).
But if we take a soccer ball, neither of the above two parameters is 
`natural'. 
It is not by chance that the FIFA laws establish the ball size by its 
circumference [``{\it of not more than 70 cm 
 (28 inches) and not less than 68 cm (27 inches)}''\,\cite{FIFA}],
 for it can be easily checked with a tape-measure. 

The situation is a bit more complicated for a sphere as large as the Earth,
and on which we live.
It is a matter of fact that at that time it was practically 
impossible to make 
immediate measurements of {\it any} of the lengths related to the dimensions
of Earth.
 One could only perform
local measurements and extend the results 
to the quantity of interest, assuming a geometrical model of Earth.
However, once a geometrical model is defined, 
it becomes of practical irrelevance which parameter is 
considered as unit, that could be the meridian, the equator or
the distance between poles and center of 
Earth.\footnote{This remains true also if an elliptic, 
rather than spherical model, is considered, though two
parameters have to be taken into account, instead then just one.
One might argue that the meridian has the advantage that 
it is bound only to the assumed rotational symmetry of Earth, 
and not to its particular shape (sphere, ellipse, 
or even something more complicate). But, apart from the fact
the second commission report~\cite{Comm2} speaks explicitly
of measuring an arc of meridian, this possibility would imply
to envisage a campaign of triangulation from the pole to the equator,
that would have been infeasible at that time.
(Remember that before 1909 the north pole
 was just an hypothetical place never reached by human beings.)
As a matter of fact, the preferred parameters of
modern geodesy to characterize 
the Earth ellipsoid (also called `spheroid')
are the equatorial radius and the flattening. In particular, the latter
is the best determined Earth parameter, given with 12 significant digits
(the value of table \ref{tab:dati_terra}
 has been rounded): $1/f=298.257\,223\,563$
(WGS84)\,\cite{Geodesy,geo_tutorial}.
}

\subsection{About the choice of the quarter of the meridian}\label{ss:quarter}
It is clear that, once 
the acad\'emiciens were bound to the decimal system, decided by the 
first commission, and the unit of length had to be related to Earth 
dimensions, the unit of length of practical use  had to be 
a small decimal sub-multiple of {\it an} Earth dimension. 
But why the quarter of the meridian, instead of the meridian itself?
The {\it Rapport}~\cite{Comm2} does not give any 
justification of the choice, 
as if all other possibilities were out of question.
And this is a bit strange. The meridian as unit of length
had no tradition at all, and there had been no discussion
about which submultiple to use.  
%Despite the defense of this choice that one finds in the official 
%documents of the Academy~\cite{Comm2} and in the literature, 
Evidence against the naturalness of the 
quarter of meridian seems to us provided 
by the fact that the 
vulgarization of the definition of the meter, as it is often
taught at school and as it is memorized by most people, is the 
forty millionth part of `something', where this `something' is
often remembered as the `equator' or the `maximum circle'.

It could be that we have nowadays a different sensitivity to the subject
(we have made a little poll among
 friends and colleagues, and our impression has been unanimously
confirmed), but we find
it hard to be rationally convinced by the arguments 
of the following kind:
\begin{quote}
{\small {\sl 
Once it has been chosen as base, will either the whole meridian 
or a sensible part 
of it be taken as a unit? 
The wholeness? Out of question! The half, that stretches from one 
pole to the other, 
may not 
be easily conceived by our mind because of the part which is located 
``below'', 
in the other hemisphere. 
This is not the case of the quarter of 
the meridian that, on the contrary, 
can be easily imagined: 
it stretches from ``one pole to the equator''. 
In the future it will be said: France opened the divider and pointed 
it on one pole and the equator, 
a sentence that will be greatly successful.
There is another reason, that is really scientific and supports the meridian: 
its quarter is the arc intersected by the right angle. 
That's right: however, why should it be considered as a further advantage? 
Simply because the right angle is considered as the natural angle, 
the angle of the vertical and the gravity. It is the unit-angle, 
the degree is nothing but its subdivision.
} (Ref.~\cite{Guedj1}, p. 55 of the Italian translation)
}
\end{quote}
What would be the alternatives?
As an exercise, we show in table \ref{tab:scelte_metro}
some possible `natural' choices of units of length based on the
dimensions of Earth, together with a reasonable decimal 
sub-multiple as practical unit. 
\begin{table}
\caption{\small Some possible choices of units of length based on the
dimensions of Earth, assumed to be spherical, 
together with a reasonable decimal 
sub-multiple as practical unit and the half period of the simple
pendulum of such practical unit. (Analogous quantities can be defined 
assuming an ellipsoid).}
\begin{center}
\begin{tabular}{lrcc}
\hline
\multicolumn{1}{c}{unit} & \multicolumn{1}{c}{decimal} 
& \multicolumn{1}{c}{practical unit} & $T/2$ \\
     &  \multicolumn{1}{c}{sub-multiple}  & \multicolumn{1}{c}{(cm)} & (s)\\
\hline
radius   & 1/10\,000\,000 & \ 64  & 0.803\\
diameter &  1/10\,000\,000 & 128 & 1.135\\
meridian &  1/100\,000\,000 & \ 40 & 0.635\\
$1/2$ meridian (pole-pole) &  1/10\,000\,000 & 200 & 1.419\\
$1/4$ meridian (pole-equator)&  1/10\,000\,000 & 100 & 1.004\\
45th parallel &  1/100\,000\,000 & \ 28 & 0.534\\
one radiant along the meridian &  1/10\,000\,000 & \ 64  & 0.803\\ 
(same as radius) & & & \\
1 degree of Earth's arc & 1/100000 & 111 & 1.057 \\
1 minute of  Earth's arc$^{(*)}$& 1/1000 & 185  & 1.367 \\ 
1 second of Earth's arc & 1/100   & \ 31 & 0.558 \\
\hline
\multicolumn{4}{l}{\small $^{(*)}$\,Equal to 1 nautical mile, 
that is 1852\,m.}\\
\end{tabular}
\end{center}
\label{tab:scelte_metro}
\end{table}
Sub-multiples of the length of the meridian, e.g. one part
over 10\,000\,000 or one part over 100\,000\,000, had led to a `meter' of
400 or 40 of `our' centimeters. The former is certainly too large, 
but the latter is quite appropriate for daily use, and, 
indeed, it falls in a range of length that is better perceived
by people (one of the criticisms to the meter is its unnaturality, at least
compared for example to the foot or even to the first standardized unit, 
the cubit).
Even the pole-to-pole arc would have yield a better practical unit,
very close to the toise.
 
We see from table \ref{tab:scelte_metro} that the 
10\,000\,000 part of the quarter of meridian is the 
closest to the length of the seconds pendulum.
So, when the French scientists proposed the new unit of length,
we think it is possible, among the many `defensible natural units', 
they chose the closest to the seconds pendulum. 
The reason could be a compromise with
the strenuous defenders of the seconds pendulum. Or it could have happened
that, since they had in mind some `cooperation' between
the new unit and ``{\it a pendulum
having a determined length}''\,\cite{Comm2}, choosing a unit close
to the well studied seconds pendulum would have 
simplified the intercalibrations.

\subsection{About the naturalness of a system of units}
The main reason to reject the seconds pendulum was 
``{\it to have a unit of length that does not depend 
on any other quantity}''\,\cite{Comm2}. Now we think 
exactly the other way around,
and prefer a system with a minimal number of units
 connected by physical laws, as it was suggested first by 
Burattini in 1675
(see Appendix B).  Besides cultural aspects, that make 
change `what is perceived as natural' with time,\footnote{With 
this respect, a
suggestion put forward in 1889 by 
 Max Planck  has been particularly influential. 
 He  proposed that  systems  of units should be based on values
assigned conventionally to certain fundamental physical constants.
The first  (partial)  realization  of Planck's idea took place in 1983
when the constancy  of the speed of light in different 
inertial frames, adopted by Albert Einstein as the grounding principle
of special relativity,
was finally used to relate the unit of length
to the unit of time.
} we find
a certain contradiction in the use of 
the naturalness concept expressed in the {\it Rapport}\,\cite{Comm2}.
Why not to extend it also to the weight unit, instead
of binding, as it is known, 
this unit to the unit of length and to the density of water?

 A similar comment applies 
to the right angle as the ``{\it natural angle}'' to justify 
the quarter of meridian
(see Guedj's quote
in subsection \ref{ss:quarter}): the right angle is certainly
the natural one for a square or a rectangle, but why should it
be natural for a circle, where there are no 
angles?
(At most, if there were an angle to be considered natural,
that would be  the radiant, as all those who use trigonometric functions
of computer scientific libraries know).

\subsection{About the reticence concerning the value of the new unit of length}
As we have stated in section \ref{lunghezza_metro}, we 
have been surprised not to find an estimate of the length
of the unit proposed in the second commission report~\cite{Comm2}.
Perhaps they were really afraid that, if the members 
of the National Assembly had known that length was determined 
within a few parts in ten thousand, then the new meridian
mission would have not been financed~\cite{Alder}.
In fact, when two years later they had to finally release 
an official number, some acad\'emiciens saw in the 
provisional meter the end of the meridian 
mission.\footnote{``{\it The new measures are being adopted 
for the commerce independent of the new measure of the earth;
so there is little need for you to push yourself too hard 
to bring your result now}'', wrote Lalande to Delambre
(cited in Ref.~\cite{Alder}%, p. 100
).}

There is another point, contained in the documents that
led to the provisional meter~\cite{BLM,luglio1793},
that has puzzled us. Why  did they not provide their
`best value', that included the best 
understanding  of Earth flattening?
As we have seen in footnote 34, the flattening correction to the meter
is about two parts in ten thousand. If applied, 
it would have changed the provisional meter from 443.44 lignes to about
443.36 lignes. A little difference, but still relevant, if
compared with the significant digits with which the provisional
meter was given. We are not arguing on the base of after-wit
arguments (the present best value of the 1/10\,000\,000 
of the quarter of meridian is 443.3975 toises,
right in the middle between the two values).
Our question is only why they did not report the `best value'
--- and they knew that 443.44 lignes was not the best
value deriving from their status of knowledge, because that value
assumed a spherical Earth, an hypothesis already 
ruled out by theoretical considerations and experimental results. 

The simplest reason could be they considered
that value good enough for practical applications
and, since the provisional meter had to be most likely revised,
it was not worth applying the little correction that was
of the same size of the expected error. But then, why to 
state that ``{\it its error does not exceed one tenth of ligne}'',
if the error due to the omitted correction counted already 
by about 0.08 lignes? (We understand that statements 
concerning errors have always to 
be taken  {\it cum grano salis}, but the expression
``{\it does not exceed}'' is quite committing).
Frankly, we do not see any plausible and consistent 
explanation, other than the somehow malicious guess,
to be taken with the benefit of 
inventory, that they wanted to have some room 
to modify the provisional meter after the end of the
new measurements, in the case the new result
would come very close to the old one. That would have
justified the expensive enterprise.

\subsection{About hidden motivations of the meridian mission}
The suspicion that the choice of the meridian based unit of length
was just a veil  for other reasons is
not a new one. Many arguments are given for example in the 
captivating Alder's {\it The measure of all things}\,\cite{Alder}. 
We are inclined to believe in the most noble cause advanced in this
book, i.e. the acad\'emiciens were mainly interested in making 
 a decided step in the understanding of the shape
of our planet. Borda's interest to see his {\it repeating 
circle} protagonist of a great enterprise seems also a 
good reason, but we see it at a second level. (It seems to us 
the anonymous' quote ``{\it Sometimes to serve the people one must 
resolve to deceive them}'', reported % at page 251 of 
in Ref.\cite{Alder},
is quite appropriate to the case.) 

However, and that was the irony of fate, the improved knowledge
of the Earth decreed that the basic premise on which the meridian
mission started was wrong: all meridians are different from
each other, and the different arcs of the same 
meridian  are unique.\footnote{Even not taking into account asperities 
of the ground, hills and mountains, 
the concept of spheroid (or solenoid), 
is just a first approximation of the Earth shape (the 'zero-th order'
approximation is the sphere). The equipotential surface of 
Earth has a complicate shape called Geoid, of which the spheroid
is a kind of best fitting curve (see e.g. Ref.\,\cite{geo_tutorial}
for an introduction, that also contains a visual representation
of the Geoid).} 
The definition of the meter had then an intrinsic uncertainty, 
such that it would have become soon or later unsuitable for a standard
of length.\footnote{That is due neither to ``{\it le difficolt\`a
ad applicare tale definizione, causate dalla forma
sferica della terra}'' ({\it the difficulties to apply such 
a definition, caused by the spherical shape of the 
Earth})~\cite{Encarta2002}, nor because 
``{\it later it was discovered that the Earth is 
not a perfect sphere}''\cite{EncartaMsn}.}
%
%\mbox{}\vspace{0.6cm}
\section*{A final reflection}
We have reasoned the plausibility of an original meter bound to the
second through the swing of the pendulum.
If our guess is true, then, despite the search 
for standards taken from nature has been the motivation for 
so many, long scientific and philosophical researches, at the
 bases of the system virtually used all over the world there 
are the throbs of our hearts. May be an original support to 
the well known statement by Jean Jacques Rousseau 
``{\it Nothing is less in our power than the heart, and 
far from commanding we are forced to obey it}.''

\section*{Acknowledgements}
It is a pleasure to thank Dino Esposito and Paolo Quintili for 
several discussions on the subject and 
for the careful reading of the manuscript.
We also thank Ken Alder, Giuseppe Barbagli, Roberto Carlin, 
Giovanni Iorio Giannoli and Daniela Monaldi for useful comments.

\newpage
\section*{Appendix A: The local `meter' and `second' in the planets of the
solar system}\label{sec:pianeti}
The well known small angle formula that gives the period $T$ 
of the simple pendulum 
(i.e. the elementary text book pendulum) as a function of its length $l$ is
\begin{eqnarray}
T & = & 2\,\pi\,\sqrt{ \frac{l}{g} }\,, \label{eq:T}
\end{eqnarray}
where $g$ is the gravitational acceleration, approximately 
equal to $9.80\,\mbox{m}/\mbox{s}^2$ on Earth.
For $l=1\,$m we get $T=2.007\,$s. Therefore, 
each swing takes 1.0035\,s, that differs from a round second
only by a few parts per thousand. Varying $g$ by $\pm 0.3\%$ (i.e.
from $9.77$ to $9.83\,\mbox{m}/\mbox{s}^2$), 
 the period changes only by  $\pm 0.15\%$.

In order to understand if there is any physical reason behind this
numerical coincidence let us try to understand the property of
Earth that mainly influences the period of the pendulum, and if there
is any simplification due to the fact that the length of the pendulum
is about $1/40\,000\,000$ of the 
meridian.\footnote{One might also think of a simplification
due to the fact the length of the pendulum is unitary. 
To readers with little
physics background we would like to make clear 
that what matters
for the pendulum period
is the length, and not the unit in which the length is expressed. 
The value of a generic quantity $Q$ is given by $Q=\{Q\}\cdot [Q]$, 
where $[Q]$ is the {\it unit of measurement} and 
$\{Q\}$ the {\it numerical value}, 
e.g. $l=1.38\,$m.  
If we change unit of measurement from system $A$ to
system $B$ ,
$\{Q\}$ and $[Q]$ change, preserving  $Q$ invariant:
$$ Q=\{Q\}_A\cdot [Q]_A = \{Q\}_B\cdot [Q]_B\,.$$
The two numerical values are then related by 
$$\{Q\}_B = \{Q\}_A \cdot \frac{[Q]_A}{[Q]_B}\,.$$
If we call M a different unit of length, such that 
$1\,M = \alpha\,$m, we get $\mbox{m/M} = 1/\alpha$. 
Therefore, a length $l = n\,$m will be 
{\it expressed} 
as $l= N\,\mbox{M} = n/\alpha\,\mbox{M}$ in the new unit
(but the {\it length} is the same: you will not grow up, 
if your height is expressed in centimeters or millimeters 
rather than in meters).
As far as Eq.~(\ref{eq:T}) is concerned, 
the numerical value of $g$ will be transformed in the same way as $l$,
namely $g=(9.8/\alpha)\,M/s^2$. As a consequence, the conversion factor 
$\alpha$ simplifies in Eq.~(\ref{eq:T}) and the period will remain
the same, as it must be. \\
On the other hand, if we consider a {\it different} length,
that has unitary numerical value in the new unit M, 
we get a {\it different} 
period of the pendulum, 
namely $T^\prime = 2\,\pi\,\sqrt{1\,\mbox{M}/(9.8/\alpha) 
\,\mbox{M}/\mbox{s}^2} = \sqrt{\alpha}\,T$. 
(For example, if the new unit of length 
is twice the meter, the half period of a simple pendulum 
of unitary length will be 1.419\,s).} 
The gross value of  $g$ depends  on mass and 
radius\footnote{The generic `radius of Earth' 
$R$ refers usually to the equatorial radius and it implies
that Earth is considered sufficiently spherical for
the purpose of the calculations.} $R$ of the Earth
with local effects due to not exact sphericity 
(see Table \ref{tab:dati_terra}), 
mass dishomogeneity and  above sea level height.
Moreover, there is a centrifugal term, null at the pole and 
maximum at the equator, due to Earth 
rotation.\footnote{At the equator, the negative centrifugal acceleration
gives a contribution to $g$ equal to 
$\Delta g_c = - v^2/R = -4\,\pi^2\,R/T^2_{rot}=-0.034\,\mbox{m}/\mbox{s}^2$
($T_{rot}=86400\,$s stands for the rotation period). 
Note that, however, in geodesy `gravitational acceleration' $g$  
indicates the overall free fall acceleration experienced by a body
and takes into account the genuine gravitational force and 
the centrifugal one. 
}
In the 
approximation of a perfect sphere, the gravitational acceleration
$g$, i.e. the gravitational force $F_G$
divided by the mass of the pendulum, is given by
\begin{eqnarray}
g & = & \frac{F_G}{m}= \frac{1}{m}\frac{G\,M\,m}{R^2} 
        =\frac{G\,M}{R^2} \,, \label{eq:g}
\end{eqnarray}
where $M=5.98\,10^{24}\,$kg  is the mass of Earth
and $G=6.67\,10^{-11}\mbox{N}\cdot\mbox{m}^2\cdot\mbox{kg}^{-2}$ 
is the gravitational constant.
Expressing the mass in terms of density $\rho$ 
and volume $V=4/3\,\pi\,R^3$, we get 
\begin{eqnarray}
g & = & \frac{4/3\,\pi\,\rho\,G\,R^3}{R^2} = \frac{4}{3}\,\pi\,\rho\,G\,R
 \,. \label{eq:g_rho} 
\end{eqnarray}
The gravitational acceleration $g$ is then proportional to the planet 
size and density.
Let us now calculate the period of a pendulum 
whose length is $1/40\,000\,000$ part of a meridian of
a spherical planet, i.e. $l_m=\alpha\, R$, 
where $\alpha=2\pi/40\,000\,000=\pi/2\times 10^{-7}$ is 
the fixed ratio between 
this  `meter' and the planet radius. The period of such a 
`planetary meter' pendulum is
\begin{eqnarray}
T(l_m) & = & 2 \pi\, \sqrt{ \frac{\alpha R}
                                 {4/3\,\pi\, \rho\, G\, R}
                          } 
         =  \frac{\pi}{\sqrt{2/3\times10^7\,\rho\, G}}\,, 
\label{eq:T_rho}  
\end{eqnarray}
and depends only on planet density, and not on planet 
mass and size separately. In particular, in the inner planets
and Earth, for which the density is approximately 5.5 g/cm$^3$, 
such a `planetary meter' pendulum would beat approximately
the second (see Tab.~\ref{tab:pianeti}). 
\begin{table}[!t]
\caption{\small Some physical data about the planets of the 
solar system, together
with the `planet meter' ($l_m=\pi/2\times 10^{-7}\,R$), 
the half period of a `planet meter' pendulum [$T(l_m,g)/2$]
and the `planet second' [$T_{rot}/86400$].
Note that Eqs.~(\ref{eq:g})--(\ref{eq:T_rho}) have been evaluated 
assuming perfect spherical and homogeneous planets, while 
the `radius' is just one half of the equatorial diameter, 
and the half period 
$T(l_m,g)/2$ is directly evaluated from nominal
value of $g$ given in this table~\cite{Pianeti}. The minus sign in the  
period indicates retrograde rotation.}
\begin{center}
{\small
\begin{tabular}{|l|crcr|rcr|}
\hline
 Planet & \multicolumn{4}{|c|}{Physical data~\cite{Pianeti}} &
          \multicolumn{3}{|c|}{One `meter' pendulum}\\
 & & & & & \multicolumn{3}{|c|}{and its period} \\
\hline 
        & Mass & \multicolumn{1}{c}{Radius} &\multicolumn{1}{c}{$\rho$}&
       \multicolumn{1}{c|}{$g$} & \multicolumn{1}{c}{$l_m$}
        & \multicolumn{1}{c}{$\frac{\,T(l_m,g)}{2}$} 
%        & \multicolumn{1}{c}{$\frac{\,T(\rho)}{2}$} 
       & \multicolumn{1}{c|}{$\frac{T_{rot}}{86400}$}  \\
       & (kg) & \multicolumn{1}{c}{(km)} & 
       \multicolumn{1}{c}{(g/cm$^3$)} & 
       \multicolumn{1}{c|}{(m/s$^2$)} & \multicolumn{1}{c}{(m)} 
        & \multicolumn{1}{c}{(s)}  
% & \multicolumn{1}{c|}{(s)} 
        & \multicolumn{1}{c|}{(s)}  \\
\hline 
Mercury & $3.30\,10^{23}$ & 2440 & 5.43   & 3.70 & 0.38 & 1.01 & 58.6\\
Venus   & $4.87\,10^{24}$ & 6052 & 5.24   & 8.89 & 0.95 & 1.03 & $-243$\\
Earth   & $5.98\,10^{24}$ & 6378 & 5.52  & 9.80 & 1.00 & 1.00 & 1.00\\
Mars    & $6.42\,10^{23}$ & 3397 & 3.93  & 3.69 & 0.53 & 1.19 & 1.03\\
Jupiter & $1.90\,10^{27}$ & 71492 & 1.33 & 23.17& 11.23 & 2.19 & 0.41\\
Saturn  & $5.68\,10^{26}$ & 60268 & 0.69 & 8.98 & 9.47 & 3.23 & 0.45\\
Uranus  & $8.68\,10^{25}$ & 25559 & 1.32 & 8.71 & 4.01 & 2.13 & 0.72\\
Neptune & $1.02\,10^{26}$ & 24766 & 1.64 & 11.03& 3.89 & 1.87 & 0.67\\
Pluto   & $1.27\,10^{22}$ & 1137 & 2.06  & 0.66 & 0.19 & 1.64 & $-6.39$\\
\hline  
\end{tabular}
}
\end{center}
\label{tab:pianeti}
\end{table}

However, the half period of this pendulum is approximately equal 
to the  $1/86400$ part of the planet rotation only for Earth
and Mars, which have approximately equal `days'. For all other
planets, the local day can be very different with respect to
the Earth one. In fact, 
the rotation speed is related to the initial angular momentum
when the planet was formed and there is no reason why it should
come out to be the same in different planets 
(Venus and Pluto are indeed retrograde, i.e. they rotate East-West). 

\section*{Appendix B: Tito Livio Burattini's catholic meter}
Among the several scientists that 
advocated the seconds pendulum
as unit of length it is worth to
emphasize the figure of Tito Livio Burattini,
an unusual and interesting personality of the 17th 
century,\footnote{Tito Livio Burattini (born 1617 in Agordo, 
Belluno, Italy
and died 1681 in Krakow, Poland) 
was an Italian Egyptologist, inventor, architect,  
scientist, instrument-maker, and traveler.
He was an extremely versatile person 
(he even designed ``flying machines''!), with interests in mathematics, 
physics, astronomy, geodesy and economics. He spent a few years in Egypt, 
where he prepared a triangulation map of this country 
(he was also an excellent cartographer), 
made measurements of many pyramids and obelisks, 
copied monuments and tried to classify them. After 
some stay in Germany, he finally settled in Krakow, where 
he served as the King's architect. There he 
 performed  optical experiments and contributed to the discovery 
of irregularities on the surface of Venus,
in collaboration with  the astronomers Stanislaw 
Pudlowsky, a former student of Galileo,  and Girolamo Pinocci.
 He became also a highly regarded maker of microscope and 
telescope lenses, sending some of them as gifts to Cardinal 
Leopold de' Medici. In 1645, he published {\it Bilancia 
Sincera}, where he proposed an improvement to the 
hydrostatic balance described by Galileo in his {\it Bilancetta}. }
and his {\it Misura Universale}, 
published in 1675~\cite{Burattini}. Apart from issues of 
priority on the proposal of the seconds pendulum as unit of length,
to which we are not interested, 
the historical relevance of Burattini's 
work resides mainly in the 
several modern concepts and nomenclature
that appeared for the first time in his book. The most relevant of them is
the idea of relating different
 units via physical quantities in order to set up a complete system
starting from the unit of time.
The sub-title in the front page of his document accounts for his 
ambitious proposal:
\begin{quote}
{\small {\sl
Treatise in which it is shown 
how in every Place
 of the World it is possible to find a UNIVERSAL MEASURE \& WEIGHT 
having no relation
 with any other MEASURE and any other WEIGHT \& anyway in every
 place they will be the
 same, and unchangeable and everlasting until the end 
of the WORLD.''}\,\footnote{``{\it Trattato nel qual si mostra come in tutti 
li Luoghi del Mondo si pu\`o trovare una MISURA, \& un PESO UNIVERSALE
 senza che habbiano relazione con niun'altra MISURA, e niun altro PESO, 
\& ad ogni modo in tutti li luoghi saranno li medesimi, 
e saranno inalterabili, 
e perpetui sin tanto che durer\`a il MONDO.}''
(The original is for the pleasure of Italian readers.)
}
}
\end{quote}
Here the word {\it universal} is used for the first time 
for a unit of measurement. 
In the 9th page 
 of his document (pages are unnumbered)
he makes the suggestion to call  
{\it metro cattolico} 
(catholic meter --- `catholic' in the sense of universal) 
a standard realized by the pendulum:
\begin{quote}
{\small \sl
So, Pendula will be the basis
 of my work, and from them I shall  first originate 
my Catholic Meter, that is
 the universal measure, as I think I have to name it in Greek, and then I
  shall originate a Catholic Weight from it.''}\,\footnote{{\it ``Dunque 
li Pendoli 
saranno la base dell'opera mia, e da quelli cavar\`o prima il mio 
Metro Cattolico, cio\`e misura universale, che cos\`\i\     mi pare di 
nominarla in lingua Greca, e poi da questa cavar\`o un Peso Cattolico.}''}
\end{quote}
Here is finally, in the 20th page, his definition of the meter:
\begin{quote}
{\small \sl
The Catholic Meter is nothing but the 
length of a Pendulum,
 whose oscillations are 3600 in a hour {\rm [...]}  as I refer to a 
free Pendulum, and not
 to those which hang from Clocks.}\,\footnote{{\it ``Il Metro 
Cattolico 
non \`e altro che la lunghezza di un Pendolo, le di cui vibrazioni 
siano 3600 in un hora {\rm [\ldots]} ch'io intendo d'un Pendolo libero, 
e non di quelli che sono attaccati agli Horologi.}''}
\end{quote}
We think Burattini would be very pleased to learn that the unit of length
of the International System differs from his meter by only half centimeter!
\newpage

%%%%%%%%%%%%%%%%%%%%%%%%%%%%%%%

\end{document}